\documentclass[preprint,12pt]{elsarticle}

\usepackage{amssymb}
\usepackage{tikz}
\usepackage[hidelinks]{hyperref}

\journal{Mathematical Social Science}

\begin{document}

\begin{frontmatter}

\title{Measuring changes in polarisation using Singular Value Decomposition of Random Dot Product Graphs}


\author[inst1]{Sage Anastasi}

\affiliation[inst1]{organization={Department of Mathematics and Statistics, Canterbury University},
            country={New Zealand}}

\author[inst2]{Giulio Valentino Dalla Riva}
\affiliation[inst2]{organization={Department of Mathematics and Statistics, Canterbury University},
            country={New Zealand}}

\begin{abstract}
In this paper we present new methods of measuring polarisation in social networks. We use Random Dot Product Graphs to embed social networks in metric spaces. Singular Value Decomposition of this social network then provides an embedded dimensionality which corresponds to the number of uncorrelated dimensions in the network.. A decrease in the optimal dimensionality for the embedding of the network graph means that the dimensions in the network are becoming more correlated, and therefore the network is becoming more polarised.

We demonstrate this method by analysing social networks such as communication interactions among New Zealand Twitter users discussing climate change issues and international social media discussions of the COP conferences. In both cases, the decreasing embedded dimensionality indicates that these networks have become more polarised over time. We also use networks generated by stochastic block models to explore how an increase of the isolation between distinct communities, or the increase of the predominance of one community over the other, in the social networks decrease the embedded dimensionality and are therefore identifiable as polarisation processes.

Corresponding author: Sage Anastasi, sage.anastasi@pg.canterbury.ac.nz, University of Canterbury, Private Bag 4800, Christchurch 8140
\end{abstract}


\end{frontmatter}

\begin{tikzpicture}[remember picture,overlay]
\hypersetup{hidelinks}
\node[anchor=north west]  at (current page.north west) 
{ \href{https://networksci.peercommunityin.org/articles/rec?id=217}{\includegraphics[height=35mm]{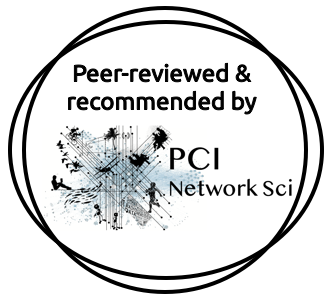}} } ;
\end{tikzpicture}

\section{Introduction}
\label{sec:sample1}

Social and political polarisation is an issue of increasing concern in New Zealand \cite{hannah_polarisation_2023}, and many other countries \cite{tucker2018social,  Hebenstreit_Germany_2022, Rodon_Spain_2022, Kozłowski_Poland_2023, Teney_Germany_2023}. While it was initially thought that New Zealand had avoided the populist takeover seen in countries such as the USA \cite{vowles_exception_2020}, radical changes in the government makeup in both 2020 and 2023 have since called this into question. It is therefore important to develop tools to measure changes in polarisation in societies so that it can be detected and countered as early as possible. In this paper we raise issues with existing polarisation measurement methods, particularly bimodality-based analyses, and propose a new method of measurement which we believe accounts for these issues.

New Zealand was considered to have escaped the development of polarisation which had developed in similar nations, particularly with the 2016 election of Donald Trump in the USA \cite{vowles_exception_2020}. Analysis of voter polarisation in NZ from 2009-2018 showed little evidence of polarisation in that period \cite{satherley_voting_2020}. Topic modelling of political party manifestos showed similarities between the manifestos of parties on the left and right \cite{orellana_manifestos_2023}, suggesting that despite their rivalries the parties are not deeply divided. Tan \cite{tan_taiwan_2020} argues that the lack of bipartisanship between NZ’s left and right indicates that the parties are polarised; however, they found that the perceived distances between left and right parties, and between those parties and voters, were stable over time, suggesting that polarisation is not increasing. Satherley et al. \cite{satherley_stability_2021} find that there is a high level of stability in the partisanship of NZ voters (i.e. they do not frequently change which party they support) in years up until 2017 and note that this creates a risk of polarisation if voters stay committed to a party that becomes increasingly extreme. In another paper, Satherley et al. \cite{satherley_identity_2020} find that in analysis of polarisation between the Labour, Green, and National parties, NZ European voters and voters of high socio-economic status exhibit more polarisation in their party preferences than other groups. Stanley et al. \cite{stanley_climate_2021} find that climate change is not a polarising issue for New Zealand, with the voting public largely accepting its existence.

This history of political stability highlights how irregular the NZ election results in 2020 and 2023 have been, with significant implications for potential polarisation. The 2020 election saw the Labour party win more than 50\% of the vote, the first outright majority since the change to Mixed Member Proportional government in 1996, and saw the right-wing ACT minor party increase their vote share from 0.5\% to 7.58\%. The covid-19 pandemic has subsequently become a focus for disinformation producers, prompting concerns about an increase in polarisation since 2020 \cite{soar_infodemic_2020, dentith_covid_2023, hannah_parliament_2022}. The 2023 election gave further new results; despite the return to the right-wing, both the NZ First and ACT parties were required for the National party to form a coalition, creating the first instance of the National party having to negotiate with two coalition partners of this size. As such, analysis of party-based polarisations based on pre-2020s electoral patterns may no longer be accurate, and developing methods that do not rely on evaluation of political parties is one way of circumventing this issue.

A significant polarisation risk in the current New Zealand landscape is the National party agreeing to support ACT’s “Treaty Principles Bill” through the early stages of the parliamentary process (without guaranteeing it will pass into law). This bill relates to the Treaty Of Waitangi, signed between the British Crown and many Māori chiefs in 1840, which gave the Crown the right to establish a government for its subjects in New Zealand, protected Māori rights to manage their own resources and affairs, and gave Māori the same rights as British citizens \cite{Orange_treaty_2015}. While the Waitangi Tribunal (the permanent commission of inquiry tasked with investigating governmental breaches of the Treaty of Waitangi) has found that Māori did not cede sovereignty by signing the Treaty \cite{waitangi_sovereignty_2014}, what this means for New Zealand in constitutional terms is complex and still evolving, especially since the country does not have a single written constitutional document \cite{Vowles_treaty_2024}. The Treaty is currently applied to law through the “Treaty Principles”, which are drawn from 50 years of case law \cite{O'Sullivan_treaty_2021}. Despite not being codified in parliamentary law, the current Treaty Principles are relatively settled \cite{ jones_treaty_2024}. The Treaty Principles Bill defines an entirely new set of Treaty Principles which contradict and overwrite existing case law, removing references to specific Māori interests. The Waitangi Tribunal  has argued that in allowing the Treaty Principles Bill to be introduced to parliament, “the Crown’s [Government’s] actions ‘threaten to disfigure or rupture’ the Crown–Māori relationship and ‘could set back the foundational relationships of Aotearoa New Zealand for decades’” \cite{waitangi_principles_2024}. A rise in general racist sentiment, as well as racist hate crimes and violence, has been documented \cite{hannah_race_2023}. As such, we think that questions about whether the polarisation may be happening along an axis other than party affiliation are particularly relevant in the current political atmosphere, since there is a risk that these policies will create a polarisation between Māori and pākehā (descendants of British settlers) which may not match the division of political party affiliations.

Polarisation scholarship in New Zealand is particularly influenced from research in the USA, due to the countries both being English-speaking Western democracies and having a long history of cooperation and political alliance; for example, New Zealand’s membership of the “Five Eyes” surveillance alliance began in 1956 and continues to the present day \cite{ Battersby_phantom_2023}.As such, it is useful to cover some key analyses of polarisation in the USA. A 2018 literature review \cite{ tucker2018social} found that the vast majority of polarisation research in the USA focussed on division between the Republicans and Democrats, at the expense of other possible polarisations such as racial segregation between whites and non-whites. Heatherington \cite{hetherington_review_2009} famously found that the Republican and Democrat parties have no overlap in their policies or voting patterns, finding a bimodal division by inspection (e.g. no overlap between the ideological groups). His focus on the Republican-Democrat axis leads him to claim that the USA was minimally polarised during the 1950s, since there was no bimodal division; however, he does not comment on other possible axes of polarisation, such as the racial segregation which was legally enforced across the country at the time. We believe that this issue of fixation on the Republican-Democrat axis at the expense of other polarisations applies to a great deal of polarisation research on the USA. Most subsequent research uses significance tests such as Hartigan’s Dip test \cite{hartigan_dip_1985} \cite{kopacheva_users_2022}. When investigating social networks, bimodality is expressed as strong in-group/out-group divisions \cite{valensise_dynamics_2022}, and is often characterised by hostility between the two groups; it is common to consider this hostility as a key sign of polarisation, in addition to the differences in policy positions between the two poles (the Republican and Democrat parties) \cite{iyengar_affect_2012, Tanesini_affect_2022}. On social media, these hostile divisions usually take the form of "echo chambers" that focus on one set of political views and exclude all others \cite{del_vicario_echo_2016, Gao_Echochamber_2023}. 

A common aspect of definitions of polarisation, first proposed by Esteban \& Ray \cite{esteban_measurement_1994}, is that the polarisation is maximised when the groups are of equal size, as well as strongly divided. Taking group size into account is intended to help tell polarisation conflicts apart from other major social conflicts, such as conflicts over wealth inequality. However, a consequence of this theoretical approach is that sets a rigid upper limit on how severe a polarisation can become if one of the groups is a small minority. We believe that there are cases where a small group is at high risk of being a target of polarisation, such as Middle Eastern and African refugees entering Europe \cite{ kopacheva_users_2022}. Similarly, the extreme racial polarisations of South African apartheid and chattel slavery in the USA both occurred with 80\%-20\% divisions between the poles (though opposite, with a white majority in the USA and minority in South Africa).  Given these issues, we are interested in finding definitions of polarisation, and tools for measuring it, that are not as affected by the size of the groups.

We have some specific concerns about the use of bimodality in detecting polarisation. The most important is that use of hypothesis tests can be used to detect when a distribution is polarised, but after the first significant result it is difficult to robustly show that any further increase in polarisation is also significant. That is to say, that if polarisation is measured at times $t$ and $t+1$ with a significant result from a hypothesis test at both, and $t+1$ has a larger test statistic than $t$, it is difficult to show that this increase itself is significant. Doing so would require finding the standard error of the difference (likely by bootstrapping) and using that to compute and evaluate a test statistic based on the observed value of the difference. Most researchers do not do this.. This means that while studies which use bimodality may detect when polarisation first occurs, we are not confident that they are able to claim that an existing polarisation is getting worse. We are therefore interested in developing measurements of polarisation that do not rely on significance testing in this way. A secondary concern is that this approach can struggle when assessing multi-party democracies that do not have a clear left- and right-wing split \cite{Röllicke_Mulitparty_2023}. Currently, the main approach to multi-party democracies is to divide all the parties into left- and right-wing blocs and then use these blocs as the two poles measured. However, in the New Zealand context, the NZ First party makes it difficult to clearly reduce the field to left and right blocs; they have formed coalition governments with both the Labour and National parties many times, and are frequently in a position to choose between forming a “left wing” or “right wing” government, so they cannot be removed from the analysis without severely misrepresenting the political field. As such, finding a method of measuring polarisation that can better handle this type of situation would be beneficial to researchers in New Zealand and similar countries.

A common preprocessing step in analyses of polarisation is to use a dimensionality reduction algorithm, such as Principal Component Analysis or Canonical Correspondence Analysis, in order to create a single dimension that can be evaluated for bimodality \cite{falkenberg_growing_2022}. These analyses do not draw any other information from the principal components, CCA weights or loadings, such as where the inflection point occurs on a scree plot of the principal components; they are simply used to discover the largest principal component or axis (i.e. the largest dimension). An advantage of this approach is that the issue driving polarisation is generated from the data, rather than researchers presuming what it is and risking choosing incorrectly, such as USA research incorrectly focussing on the Republican-Democrat axis – as we have already discussed. However, it does not entirely mitigate the risk of choosing the wrong dimension, since it presumes that polarisation is happening along the first dimension, and it therefore will still miss cases in which polarisation is happening along one of the other dimensions and has not yet grown severe enough to become the largest dimension. Furthermore, if more than one dimension is required to describe the data (i.e. the elbow of the PCA scree plot occurs at a higher value than 1) then results that find that it is bimodal along the first principal component are not accounting for the fact that the data may not be separated at all along the second, third etc. principal components. When analysing a social network, each principal component reflects a source of connections between individuals; as such, even if the first dimension has a bimodal distribution, fully polarised separation of the network is prevented by connections created in other dimensions. This creates a serious risk of detecting polarisation when it is not actually occurring that we believe is not adequately addressed by many studies which use this method. Due to these issues we are interested in developing methods which could capture polarisation occurring in the network before it becomes severe enough to become the largest dimension, and which do not create false positives by eliminating key dimensions .

In some cases, clustering algorithms are used to measure polarisation instead of bimodality. This usually incorporates the same preprocessing step of using PCA or CCA to reduce the data to a small number of dimensions of interest, such as reducing the data to two dimensions to allow visual inspections of concentrations. However, this has similar issues to bimodality testing, as it is difficult to interpret whether the data is polarised when there are more than two clusters. Methods for detecting polarisation through clustering focus on measuring distances between the clusters. A large distance between the clusters can indicate polarisation between the groups if the clusters also have low internal variation \cite{schubert_stop_2022}. This means that in theory it is possible for many groups to be polarised against each other; however, we believe there are reasons to restrict the definition of polarisation to only the two group case. In order for all n groups to be the same distance apart (i.e. to dislike all other groups the same amount), there must be at least n-1 dimensions in the space that the group clusters occupy.  For example, while two groups A and B could reach polarisation in a 1-dimensional space, the introduction of a third group C would make it impossible for all three groups to be polarised as it would not be possible for A, B and C to all be equidistant from each other. As such, a second dimension of measurement must be added in order to allow equidistant separation between three groups, a third dimension for four groups, etc. Each added dimension increases the total amount of space in between the polarised positions, which creates more possible ways for the groups to be in positions other than equidistance (i.e. more ways for them to be “not polarised”). Since the likelihood of all groups becoming equidistant decreases as dimensions of measurement are added, and since it is impossible for n groups to all be equidistant in a space with fewer than n-1 dimensions, we choose to restrict the definition of “polarisation” to the two group case.

In order to comprehensively address the issue of correct issue selection, Baldassarri and Gelman \cite{Baldassarri_partisans_2008} propose a definition of polarisation as increasing correlation in the ideological space, i.e. that pairs of ideological issues asked about in the American National Election Study become more correlated as a result of polarisation, reducing political pluralism and restricting possible ideological opinions, until maximum correlation is reached and an oppositional binary is created. This polarisation stands in comparison to an integrated, non-polarised society which "is not a society in which conflict is absent, but rather one in which conflict expresses itself through nonencompassing interests and identities". Using this definition, they did not find that there was increasing correlation in the ideological field of US-American voters pre-2004. However, subsequent research \cite{Kozlowski_alignment_2021} using the same methods found that polarisation rapidly increased between 2004 and 2016. They noted that the increase in polarisation was strongest in the domains of economics and civil rights issues, rather than in the domain of moral issues that the "culture war" framing of polarisation may suggest. Similarly, DellaPosta \cite{DellaPosta_oilspill_2020} conceptualises polarisation as similar to an oil spill, with the increasing correlation in ideological positions spreading polarisation to previously apolitical members of society. The article analyses how the "belief network" of US-American politics has changed over time, concluding that the network has developed clusters which have reduced the prevalence of cross-cutting ideological positions; this means that pluralism has decreased and polarisation has increased.

Incorporating post-structuralist political theory allows us to expand on this understanding of polarisation as correlation. Ernesto Laclau and Chantal Mouffe \cite{laclau_hegemony_1985} begin from the same understanding of pluralism and polarisation as scholars such as Baldassarri and Gelman, but they develop this concept beyond just the ideological field. They argue that "In a colonized country, the presence of the dominant power is every day made evident through a variety of contents: differences of dress, of language, of skin colour, of customs [...] the colonizer is discursively constructed as the anti-colonized." (p. 128). This is to say that it is possible for elements of a society's culture that are not explicitly political to become associated with ideologies and drawn into the correlation, further reducing opportunities for political pluralism. Two poles are constructed that are mutually exclusive and have nothing in common, sustained by segregation in all layers of society (e.g. the South African regime of racial apartheid). Note that this is an extension of the argument we made earlier about three-group polarisations; not only are such polarisations unlikely, but the social dynamics that create polarisation prevent them from occurring, because any pair of groups has a “common enemy” of the remaining group. This “common enemy” prevents the groups from becoming completely separated from each other. Such a society would become “polarised” when the three blocs become collapsed into two fully opposed groups, such as by one of the groups being destroyed or merging with another group. In the case of South African apartheid, this required sorting ethnicities such as Chinese and Japanese into the existing polarisation through declaring them to be “coloured” or “honorary whites” (which did not have exactly the same rights as “white”, but more than “coloured”); this approach preserved the overall two-group polarisation by attaching these ethnicities to the two existing poles rather than allowing them to form a clear third group of their own \cite{Sugishita_honorary_2017, Park_apartheid_2008}. Finally, the polarisations that Laclau and Mouffe examine do not occur primarily in the division between political parties, but along fault lines such as race and ethnicity, economic class, the urban-rural divide, and the division between coloniser and colonised. We therefore think that it would be beneficial to develop methods that do not presuppose that political polarisation is happening between the left- and right-wing political blocs, as this may not capture the underlying polarisation correctly.

In this paper we present a novel method of measuring polarisation that follows from both Laclau  Mouffe \cite{laclau_hegemony_1985}, and Baldassarri and Gelman \cite{Baldassarri_partisans_2008}, namely using Singular Value Decomposition to determine the embedded dimensionality of social networks. We draw from similar uses of this method in ecology to find the number of niches in an ecological community \cite{Stouffer_plant_2021}. Since the embedded dimensionality corresponds to the number of uncorrelated singular values, this corresponds to how much correlation there is in the structure of the network. Using social networks allows us to capture interactions between people that are not explicitly political — being neighbours, sharing a workplace, etc — but which become politicised and segregated during extreme polarisation. As such, we are able to determine whether correlation is increasing not just among possible political positions, but whether it is increasing among social determinants of interaction as well. Our method gives a value that corresponds to the network's capacity for complexity, and is inversely related to its level of polarisation. Incorporating these additional layers of a society into the analysis should make it easier to detect whether polarisation is occurring.

\section{Methods}
For our new method of measuring polarisation, we define the mathematics of Random Dot Product Graphs and how to use them to infer the properties of a given network graph that indicate whether it is becoming more or less polarised. We also discuss which method we use for determining the optimal embedded dimensionality of the network graph, since this is the parameter that indicates whether its polarisation is changing. Finally, we add SVD entropy to our analysis to evaluate it as another potential method for measuring the complexity of a network graph.
\subsection{Polarisation}

We define a \textit{process of polarisation} as the loss of dimensionality of a graph observed over time. Namely, we find the optimal RDPG embedding dimension $\hat{d}$ at multiple time points. If $\hat{d}$ decreases over time, then we argue that the network has become more polarised during that time. Since $\hat{d}$ is a measurement of the number of uncorrelated dimensions in the network, using it to measure polarisation is in keeping with Baldassari and Gelman’s definition of polarisation as increasing correlation of political positions \cite{ Baldassarri_partisans_2008}. A decrease in $\hat{d}$ means a decrease in the number of uncorrelated dimensions in the network, i.e. an increase in the number of correlated dimensions and therefore an increase in polarisation. In addition to $\hat{d}$, we compute the SVD entropy of the network to see whether this is a suitable tool for measuring polarisation. Since non-polarised networks have more possible configurations than non-polarised networks, i.e. they are less ordered, we are interested in whether measuring their entropy is another potential method of capturing changes in polarisation. 

\subsection{Network modelling}

We model social networks as Random Dot Product Graphs (RDPGs) \cite{CareyPriebeSurvey}. We chose this model instead of other graph embeddings because there is a data-driven method for establishing their optimal embedding dimension (see below), which is robust to changes in the network’s size.
\subsubsection{Generating a graph as an RDPG model}
In the most general, directed, case under the RDPG model, each node $i \in \{1,...,N\}$ in a graph $G$ is associated with two vectors of traits, $L_i$ and $R_i$,  that give the node position in a pair of metric spaces, $Left$ and $Right$.  All the node positions are such that the dot products $$L_i \cdot R_j$$ are in the interval [0,1] for each $i, j \in \{1,...,N\}$.. Then,the probability that an edge from node $i$ to node $j$ exists is defined by the proximity of $L_i$ and $R_j$, namely by the dot product $$L_i \cdot R_j = \mathbb{P}(i \rightarrow j) \, .$$
In other words, the position of a node in $Left$ describe its outgoing edge topology, and the position of a node in $Right$ describe its incoming edge topology.
In general, we define two matrices $L$ and $R$, such that the row $i$ of $L$ is $L_i$ and the column $i$ of $R$ is the vector $R_i$.The edge $i$ -> $j$ of the graph $G$ is hence drawn with independent probability given by the entry $(i,j)$ of the matrix product $L R$. We call the couple $(L,R)$ the \textit{RDPG embedding} of $G$.
\subsubsection{Inferring an RDPG model}
When we observe a graph, we assume that it is generated by an RDPG model, but we do not know the specific $L_i$ and $R_i$ of the model, and therefore the goal of our inference task is to estimate the position of the nodes in the latent spaces, given the interaction structure of the network. For a fixed dimension $k$ of the two latent spaces, this is achieved by a $k$-truncated Singular Value Decomposition as follows (full description of SVD can be found in Noble \cite{noble_algebra_1969}.

Let $A$ be the adjacency matrix of $G$. Let $A = U \Sigma V'$ be a singular value decomposition of A, so that $U$ and $V'$ are orthogonal matrices, and $\Sigma$ is the diagonal matrix whose $i$-th entry is the $i$-th singular value of $A$ (sorted in decreasing order). Notice that in general $U$ and $V'$ are only identifiable up to orthogonal transformations (any rotation of them would keep the dot product constant, so they would determine the same graph). We denote ${\Sigma_k}$ the truncation of ${\Sigma}$ to the first $k$ diagonal elements, giving a matrix with $k$ rows and $k$ columns. Denoting $M\vert_k$ the truncation of a matrix $M$ to its first $k$ columns, for any $k$, the two matrices $\hat{L} = U\vert_k \sqrt{\Sigma_k} $ and $\hat{R} = \sqrt{\Sigma_k} \left(V\vert_k\right) '$ determine a rank-$k$ optimal approximations of $A$. That is, $\hat{L} \hat{R} = \hat{A}$ minimizes the Frobenius distance to $A$ between all the rank-$k$ matrices. $\hat{L}$ has $N$ rows and $k$ columns, and $\hat{R}$ has $k$ rows and $N$ columns 

In the undirected case, $\hat{L} = \hat{R}$ so that $\hat{L}_i \cdot \hat{R}_j = \hat{R}_i \cdot \hat{L}_j$ and the probabilities of interaction are symmetric.

\subsubsection*{Network representation}
We represent the conversation happening on a social platform (Twitter, Facebook, Instagram, etc.) as a network. Each user that took part in the conversation is mapped to a node. We add an edge between two nodes if the two respective users have communicated in the time window of the observation. Depending on the chosen social platform considered and the specific research question, a communication can be given by a reply, a mention ("tagging" them in a post), a quote, a repost/retweet/share, or a set of these. These networks can be directed (as is more common) or undirected. Here, we consider communication networks only as unweighted graphs, although the generalization to weighted graphs doesn't present any fundamental challenges.

At the time the data was collected in 2021, their terms of service required researchers to only retain the tweed IDs rather than all the data. This was designed to allow researchers to reconstruct the networks for re-analysis as needed without creating data security and copyright problems. However, after Elon Musk’s takeover in 2022, Twitter’s download access was changed to make it extremely expensive – sometimes completely impossible – to download these networks.

\subsubsection*{Embedding dimension}

We define the \textit{dimension} of a communication network as the \textit{optimal} choice of $d$ for the RDPG embedding of the network and denote it $\hat{d}$.An a-priori optimal choice for $\hat{d}$ can be obtained from, $\Sigma$, the sorted sequence of singular values of the network's adjacency matrix $A$. Various methods exist. Here we adopt the elbow method presented in \cite{zhu2006automatic}. The elbow method identifies the most likely change point in the sequence of values of $\Sigma_k$ by sequentially fitting two Gaussian distributions with independent mean, and equal variance. One Gaussian distribution is fitted to the largest $d$ singular values, and the other to the smallest $k-d$. Then, the optimal $\hat{d}$ is the value of $d$ that maximise the sum of the log-likelihoods of the two distributions. Notice that $\hat{d}$ is robust to network size, i.e. its value is determined by the complexity of the network rather than the size of the network \cite{CareyPriebeSurvey}.

\subsubsection*{SVD Entropy}

Given a network, we can assess its graph complexity by computing its SVD entropy.  We choose this instead of other entropy measures because it is based on $\Sigma$, meaning that it will be easier to compare it to the results for the network’s embedded dimensionality than if we had chosen an entropy not based in $\Sigma$. It has also previously been used in ecology to measure the complexity of ecological networks \cite{Strydom_entropy_2021}, so we think it would be useful to test whether it works for the complexity of social networks. A network has higher SVD entropy when many of its singular vectors are highly important for its structure, meaning that the network cannot be efficiently compressed. This is commonly read as an indication of high network complexity \cite{Gu_entropy_2016}, and it is related (although not in a linear nor straightforward way) to its dimension. We normalise the SVD entropy using Pielou's evenness \cite{Pielou_ecological_1975}, so that the results do not depend on the network size.

In particular, let $\Sigma$ be the sequence of singular values of a network's adjacency matrix $A$. The \textit{nuclear norm} of $A$ is given by the sum total of $\Sigma$ (that is, the sum of all singular values).
We define the normalized values $s_i = \frac{\sigma_i}{\|A\|_*}$ where $\|A\|_*$ is the Frobenius norm of $A$, and therefore $i\in \{1,....N\}$ and $\sigma_i$ is the $i$-th singular value.

Then, the (Pielou normalised) SVD entropy of a graph $G$ is given by
$$J = - \ln(N)^{-1} \times \sum_{i=1}^Ns_i\ln(s_i)$$
where the sum term in the definition is, indeed, an entropy.

\subsection{Data, script and code availability}

Our analysis was conducted using the Julia programming language. Scripts, data, and code can be downloaded from https://doi.org/10.5281/zenodo.15180262 and run on a computer that has Julia installed. A README file is included in the repository \cite{Anastasi_repository}.

\section{Results}
We apply our computational framework for polarisation to three different data sets: two from Twitter, and one consisting of simulated interaction networks. These have been chosen in order to show that our method works well, and to explore some common beliefs about polarisation in a new way.

\subsection{Climate discussion in New Zealand Twitter}

We obtained 12939 tweets by querying Twitter's Academic API v2.0 for keywords related to climate change: climate, pollution, agw (anthropic global warming), CO2, and carbon. Some keywords used in other research, such as “COP 2x” for the international climate change conference, did not return many results when searched on New Zealand tweets. We restricted our search to original tweets geographically tagged as being from New Zealand, and tweets published after 2017, in order to investigate whether climate discussions in New Zealand were becoming more polarised. 

We divided the data into two time windows that corresponded to equally sized, large networks: between 2017 and 2020, and 2020 to 2023. For each time frame, we built a network by considering each user (identified by their unique IDs) as a node, and any mention, reply, or quote tweet between two users as an edge. Retweets were excluded in order to maintain focus on the New Zealand network and prevent the network expanding to international discussions of climate change. There were 6767 tweets in the 2017-2020 network and 6172 in the 2020-2023 network; as such, the network sizes given are based on the number of nodes unless otherwise stated. We analysed the two networks independently.

\begin{table}[hbt!]
\begin{tabular}{|c||c|c|c|c|}
\hline
\textbf{Time Period} & \textbf{Tweets} & \textbf{Users} & \textbf{Edges} & \textbf{Giant Component} \\
\hline
2017-2020 & 6767 & 1402 & 2258 & 914 \\
\hline
2020-2023 & 6172 & 1217 & 1997 & 1013 \\
\hline
\end{tabular}
\caption{Characteristics of NZ Climate Change conversation networks.}
\end{table}

We computed the results both in the original networks and in the original giant component of the network (the table of pointwise estimates below). We also computed results for 1000 bootstrapped samples of networks where we sampled (with repetition) the same amount of nodes of the original giant components of the graphs (the box graphs below). This bootstrapped sampling would allow to detect possible effect of sample limitations in the original graph – we did not find any notable sample limitations.
\begin{figure} [hbt!]
\includegraphics[width=\textwidth]{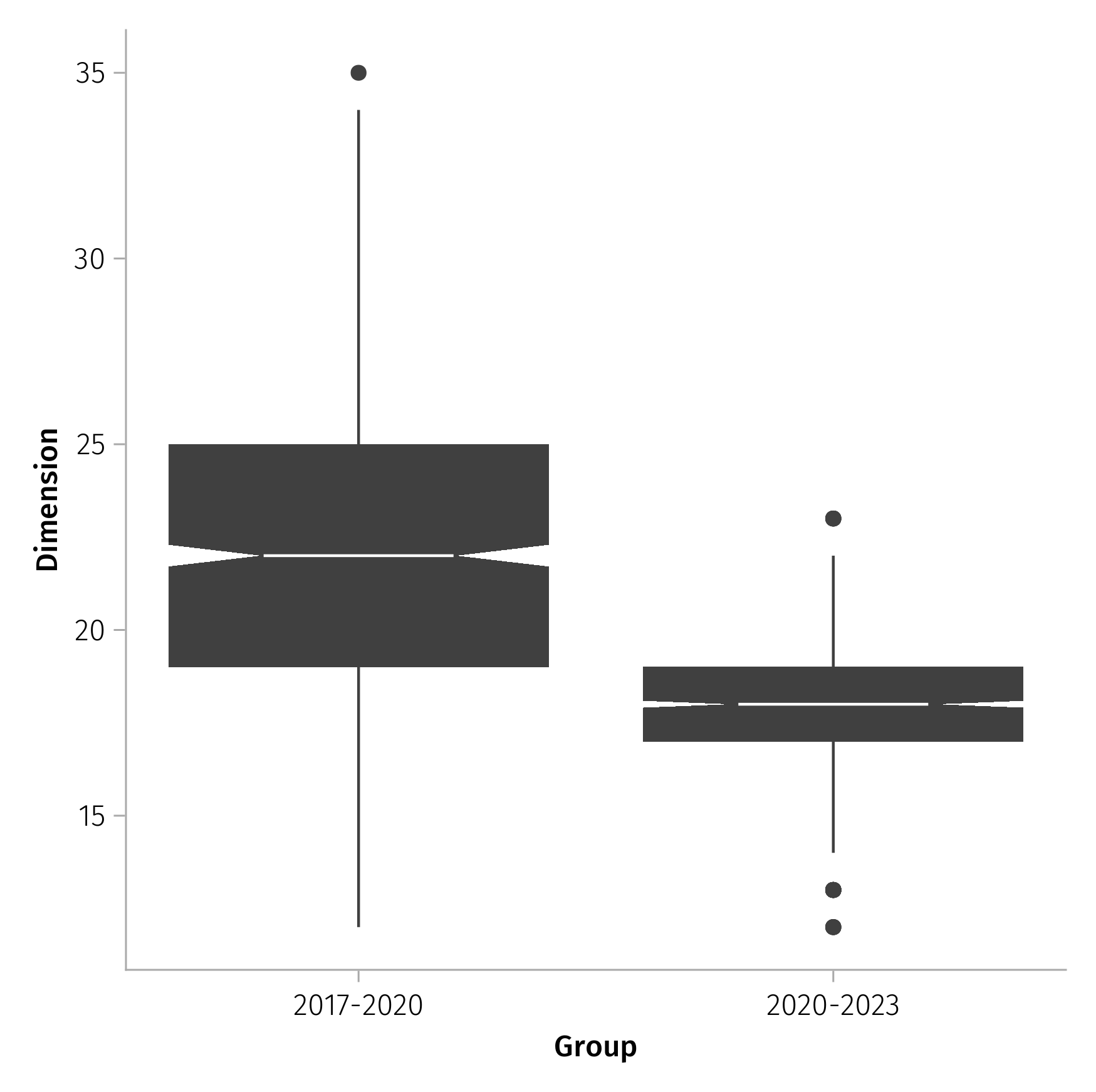}
\caption{Plot comparing $\hat{d}$ of NZ climate change tweets in 2017-2020 to $\hat{d}$ in 2020-2023. $\hat{d}$ has decreased in the second time period, which may indicate that the network is more polarised than in the first time period.} \label{fig1}
\end{figure} 

\begin{figure} [hbt!]
\includegraphics[width=\textwidth]{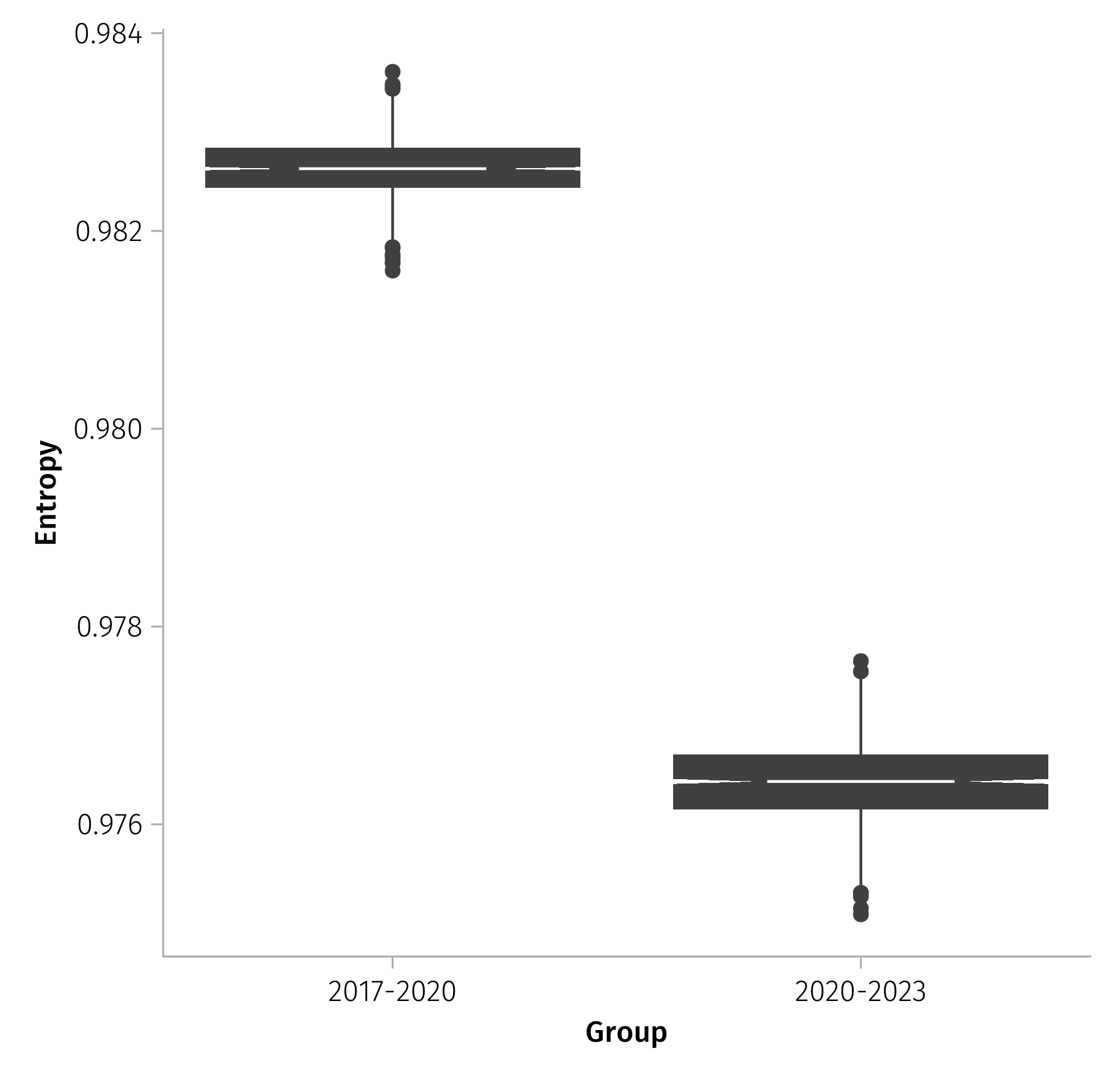}
\caption{Plot comparing the SVD entropy of NZ climate change tweets in 2017-2020 to the entropy in 2020-2023. The entropy has decreased in the second time period, which may indicate that the network is more polarised than in the first time period } \label{fig2}
\end{figure}

\begin{table} [hbt!]
\begin{tabular}{ |c||c|c|c|c|  }
 \hline
 Year&Dimension&Dimension GC&Entropy&Entropy GC\\
 \hline
 2017-2020   & 39    &39&0.980& 0.980\\
 2020-2023   & 27    &24&0.974& 0.974\\

\hline
\end{tabular}
\caption{Pointwise estimates of network dimensionality and entropy.}
\end{table}

We find that $\hat{d}$ for the NZ Twitter discussion of climate change has decreased in the 2020-2023 network compared to the 2017-2020 network, indicating that the network was more polarised in the second time period than in the first. Similarly, the SVD entropy of the network is lower in the 2020-2023 period than in the 2017-2020 period. These results  suggest that the complexity of the network decreased between these two time periods. This would support the argument that the political positions held by users are becoming narrower over time, which matches Baldassarri and Gelman's definition of polarisation.

\subsection{COP discussion in Twitter}

Falkenberg et al. collected a very large corpus of tweets by querying Twitter's Academic API v2.0 for tweets mentioning "COP2x" where x was an integer between 0 and 6 (inclusive) \cite{falkenberg_growing_2022}. We reconstructed the same networks based on the tweet IDs given in the data released with the paper. Unfortunately due to changes in Twitter’s Acadimic API in 2022, this reconstruction process is now at best prohibitively expensive, and at worst impossible. 
They restricted their search to tweets in English, and covered the COP from 20 to 26 (years 2014 to 2022, with 2020 and 2021 skipped due to the Covid-19 pandemic). Their adjacency matrix was constructed based on whether a user $i$ retweeted tweets from a  political influencer $j$. Their focus was whether there was noticeable division among users based on whether they were spreading true information about climate change or disinformation from climate change denialist influencers. The smallest network was 87,780 users and the largest was 7,809,303. More information about the networks can be found on p.39 of the supplementary materials in \cite{falkenberg_growing_2022}.To test for polarisation using our framework, we built a network for each year’s COP using the same data by considering each user (identified by their unique IDs) as a node, and any mention, reply, or quote tweet between two users as an edge. Unlike Falkenberg et al. we do not make a distinction between influencers and regular users, in order to capture the effects that interaction between normal users has on the dimensionality of the network. The network for each year was analysed independently. We then graphed $\hat{d}$ and the SVD entropy for each year in order to show how they, and therefore the polarisation of networks that discussed the COP conferences, changed over time.
\begin{figure} [hbt!]
\includegraphics[width=\textwidth]{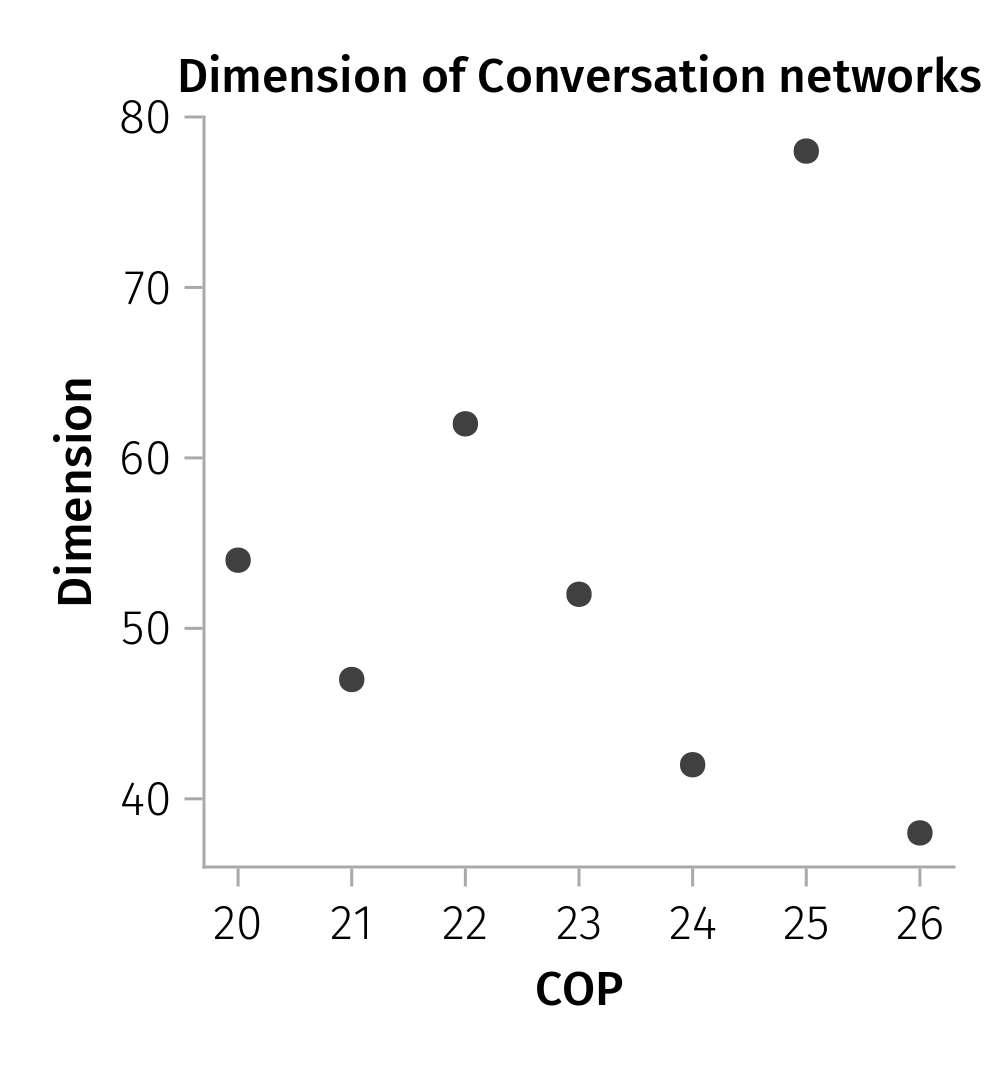}
\caption{Plot comparing $\hat{d}$ of tweets for COP20-COP26.} \label{fig12}
\end{figure} 

\begin{figure} [hbt!]
\includegraphics[width=\textwidth]{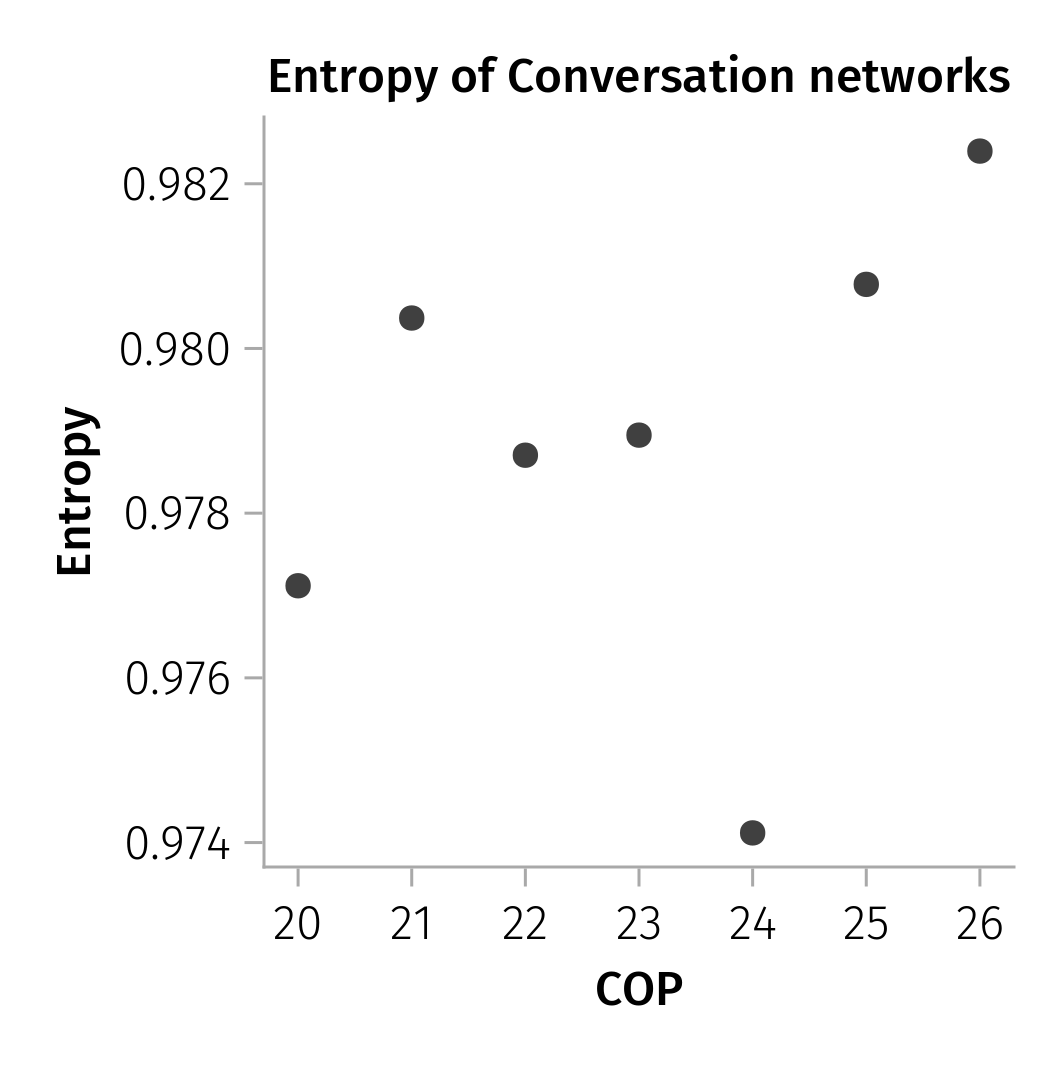}
\caption{Plot comparing the network SVD entropy of tweets for COP20-COP26.} \label{fig3}
\end{figure}

\begin{table} [hbt!]
\caption{Pointwise estimates of $\hat{d}$ and entropy based on the first 100 SVD values.}
\begin{tabular}{ |c||c|c|  }
 \hline
COP&Dimension&Entropy\\
 \hline
20&14&0.980\\
21&7&0.978\\
22&2&0.976\\
23&3&0.976\\
24&9&0.975\\
25&3&0.979\\
26&2&0.975\\

\hline
\end{tabular}
\end{table}

\begin{table} [hbt!]
\caption{Pointwise estimates of $\hat{d}$ and entropy based on the first 1000 SVD values.}
\begin{tabular}{ |c||c|c|  }
 \hline
COP&Dimension&Entropy\\
 \hline
20&54&0.977\\
21&47&0.980\\
22&62&0.979\\
23&52&0.979\\
24&42&0.974\\
25&78&0.981\\
26&38&0.982\\

\hline
\end{tabular}
\end{table}

We found that $\hat{d}$ for the network of Twitter users discussing the COP conference has been decreasing over time, though the decrease was not linear. Unexpectedly, the SVD entropy of the network did not decrease in this way, and instead it was at its highest in 2022 even though $\hat{d}$ was at its lowest.
To make sure that our results were not being driven by changes in the network size, we tested the Pearson correlation of the $\hat{d}$ values with the number of users in the network. For $\hat{d}$ based on the first 100 singular values the correlation had a p-value of 0.34 and for $\hat{d}$ based on the first 1,000 singular values the correlation had a p-value of 0.3542. We are therefore confident that changes in the size of the networks are not affecting our results. Interestingly, regardless of whether calculation of $\hat{d}$ was restricted to 100 or 1,000 singular values, COP20 had a $\hat{d}$ value the same as or greater than other years despite having a much smaller network, and COP26 had the lowest $\hat{d}$ despite having the largest network.
Interestingly, Falkenberg et al. expected to find polarisation during COP21, due to the signing of the Paris Agreement at COP21 \cite{falkenberg_growing_2022}. Their Hartigan's Dip Test for COP21 returned a significant result (p = 0.003), but they go on to claim that COP21 was not polarised despite this result. In our data, COP21 has a lower $\hat{d}$ than the years before or after. It may be possible that the network as a whole became more polarised, which is captured by our data, but this effect had not yet occurred among the "influencers" that Falkenberg et al selected. Our results support Falkenberg et al's suggestion that the increase in polarisation they observed was due to an increase in the prominence of anti-climate and generally far-right influencers on Twitter, since COP26 was the conference with the lowest $\hat{d}$.

\subsection{Synthetic Data}
We wanted to explore what happens when common understandings of polarisation are evaluated using our approach. This both allows us to test these common understandings from new angles in order to see whether they hold up, and to find out whether there are specific cases where our methods may need to have corrections applied. We use stochastic block models for these experiments because they are straightforward to model using RDPGs while also being a good match for the experiments we wish to run. We focus on experimental computation of the statistical properties rather than analytical computation because the stochastic block models are the platform for our experiments rather than objects of interest in and of themselves. 

\subsubsection{Engagement Between Two Groups}

The first common understanding we investigate is whether polarisation decreases as engagement between two groups increases (i.e. as echo chambers are removed). Understanding the effectiveness of this anti-polarisation strategy is important for implementing it in real-world social networks. We simulated a stochastic block network of 1000 nodes, split into two equally sized groups. We varied the probability of each node forming a connection within its group, and varied the probabilities of each node forming connections. In particular, we simulated a networks with in-group link probability of 0.3 to 0.45 with steps of 0.05. and between-group probabilities of 0.1, 0.05, and 0.01. Each combination of in-and between-group probabilities was simulated 100 times.

\begin{figure}
\includegraphics[width=\textwidth]{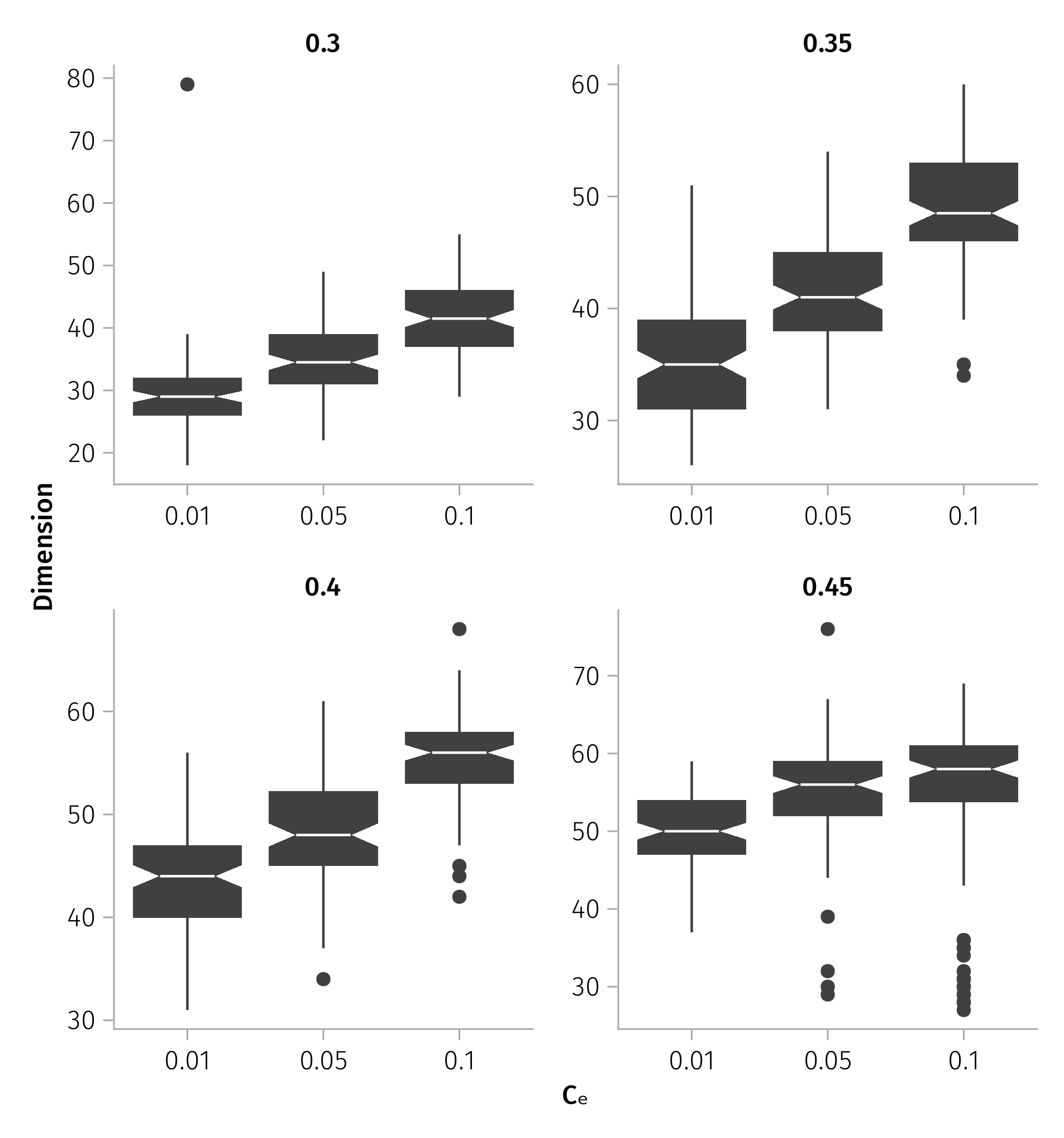}
\caption{$\hat{d}$ of the stochastic blockmodel as the link probabilities are changed.} \label{fig4}
\end{figure}

\begin{figure}
\includegraphics[width=\textwidth]{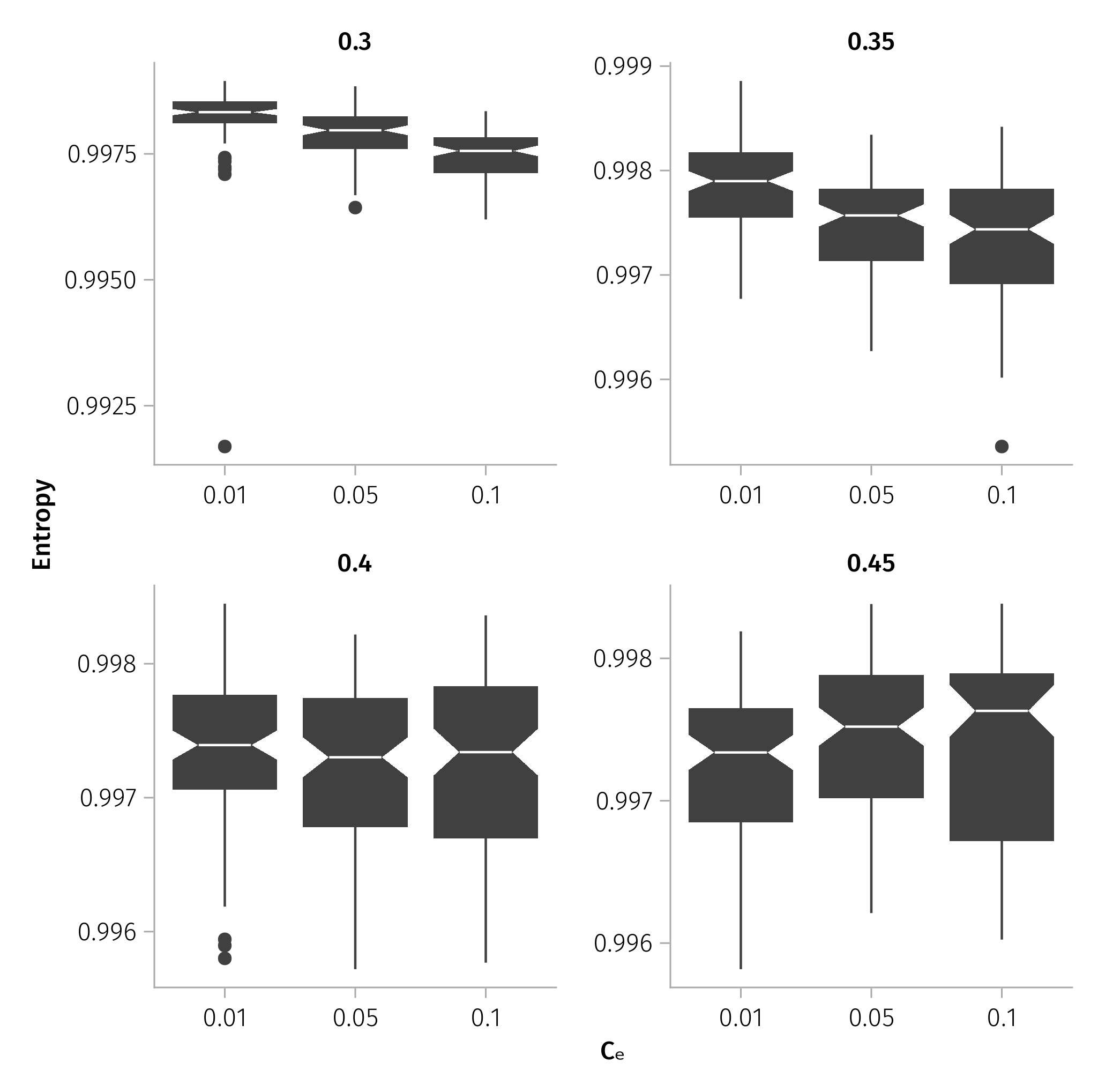}
\caption{SVD entropy of the stochastic blockmodel as the link probabilities are changed.} \label{fig5}
\end{figure}
As expected, increasing the chance of connection between the two blocks increases $\hat{d}$ (and therefore decreases the polarisation of the network). The effect was consistent across all in-group link probabilities tested. This indicates that a potential social strategy to decrease polarsation could include facilitating the creation of connections between different groups. 

For cases where the in-group link probability was lower, the SVD entropy decreased as the out-group link probability increased. In cases where the in-group link probability was higher, the entropy remained consistent or increased as the out-group link probability increased. As such, SVD entropy may be a less reliable indicator of polarisation than $\hat{d}$.

\subsubsection{One Group Becoming Larger}

The second common understanding we tested is that polarisation decreases as one of the groups becomes much bigger than the other. This has implications for real-world cases of polarisation where one group is much larger than the other, e.g. ethnic minorities or refugee communities.

We simulated a stochastic block network of 1000 nodes, split into two groups. We fixed probabilities of in-group linking between 0.3 and 0.45 with steps of 0.05, and fixed the probability of between-groups linking at 0.05. We varied the sizes of the two groups, progressively increasing the size of one group from 0.5 of the full network to 0.2, 0.1, and finally 0.01. Each combination of block size and in-group linking probabilities was simulated 100 times.

\begin{figure}
\includegraphics[width=\textwidth]{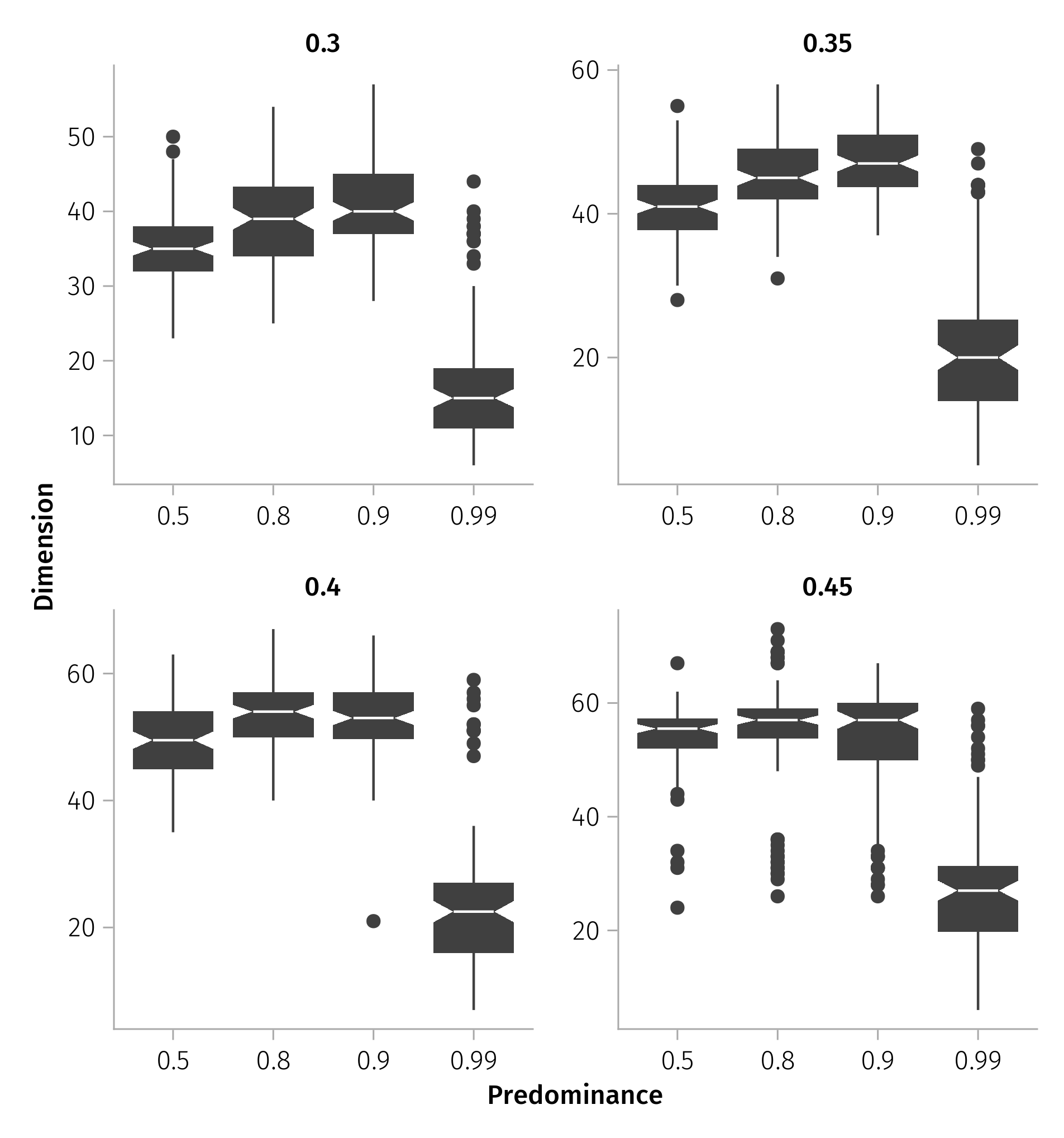}
\caption{$\hat{d}$ of the stochastic blockmodel as the sizes of the blocks are changed.} \label{fig6}
\end{figure}

\begin{figure}
\includegraphics[width=\textwidth]{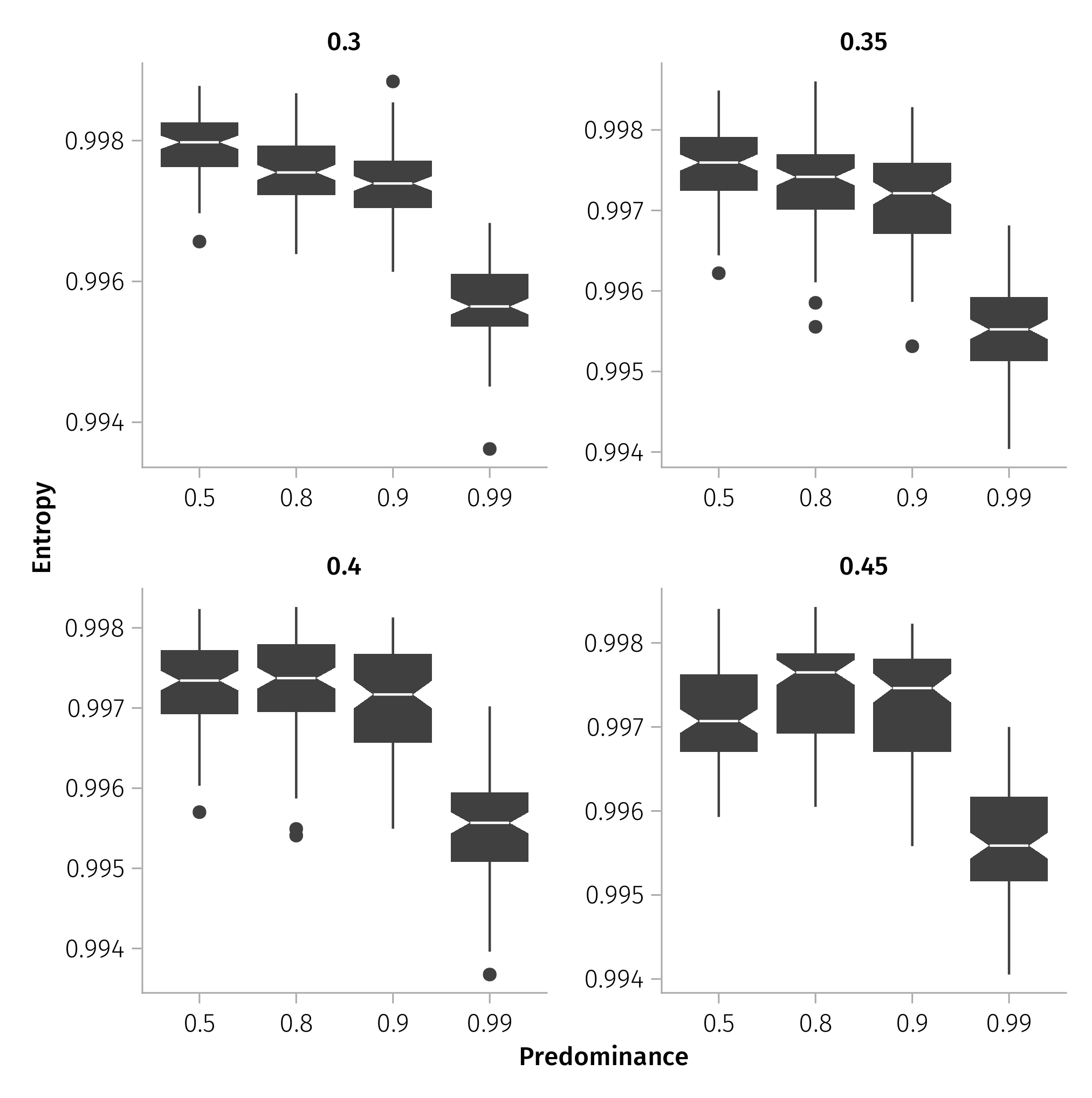}
\caption{SVD entropy of the stochastic blockmodel as the sizes of the blocks are changed.} \label{fig7}
\end{figure}

We found that $\hat{d}$ increases slightly as one group becomes predominant in the network, but decreases strongly when one group is much larger than the other. This effect was consistent across all in-group link probabilities tested. The SVD entropy of the network also strongly decreased when one group was much larger than the other (99 to 1), but did not exhibit the same behaviour as $\hat{d}$ when the group was only starting to become predominant (80 to 20, and 90 to 10). At low in-group link probabilities, the SVD entropy decreased as one group became predominant; at higher in-group link probabilities, the entropy either remained stable or increased slowly as one group became predominant. 

It is important to note that this low $\hat{d}$ is because at the greatest group size difference, the smallest group only contains 10 nodes. This means that there is a risk of the smaller group becoming disconnected entirely from the larger group, and that even at its greatest likelihood of connecting the smaller group is only sparsely connected to the larger group. While it would be possible to redesign the experiment to focus on edge density rather than only connection probability, we think this would correspond conceptually to a change in how people in the network were making connections with each other, and we wished to keep this aspect the same in our experiment rather than varying it. We believe that the sparsity in our experiment is a useful feature, as some noted polarisations have one group that is much smaller than the other (ethnic minorities, refugee populations, and LGBT+ communities being examples); in these polarisations, effects created by one of the groups being very small are relevant, as the current understanding of polarisation would suggest that there is a mathematical upper limit to how severe polarisation against a small group can become – we suspect that this is not accurate.

It is possible that our experiment did not decrease the group size far enough to trigger the effect expected by Esteban and Ray \cite{esteban_measurement_1994}. However, it does demonstrate that polarisation does not linearly decrease as one group becomes predominant, as was expected, and that the behaviour of the stochastic block model is more complicated. 

\section{Conclusions}

We have demonstrated a novel method for measuring polarisation using the embedded dimensionality $\hat{d}$ of random dot-product graphs. This is a reliable and straightforward implementation of the correlation-based approach to polarisation suggested by Baldassarri and Gelman \cite{Baldassarri_partisans_2008}, and supported with observations from political scientists Laclau and Mouffe \cite{laclau_hegemony_1985}. Our method captured the presence of polarisation in all the scenarios where it was expected and had been found by other researchers, in both simulated data and real social media networks. In particular, the ability to observe the changes in $\hat{d}$ over time in data from the COP conferences provided a useful demonstration of how this method can be applied to longitudinal data. The RDPG approach also allows us to easily see that the process of polarisation is occurring in a network over time, through its embedded dimensionality reducing, rather than relying on a binary test of whether the network was polarised or non-polarised. 

Another advantage of the RDPG approach is that it is computationally light; the main bottleneck is the computation of the first singular values of a large matrix, but this is well known in computer science literature and has already been strongly optimised. We found that the SVD was feasible even when used on networks with millions of nodes. Bimodality-based methods typically use SVD or correspondence analysis to determine the dimension they will test for bimodality, so our approach is at least as efficient.

Our approach is also highly interpretable, without forcing the latent ideological distributions into an artificially unidimensional space. Rather than creating a unidimensional space and then interpreting its political meaning (such as pro- and anti-climate, or left- and right-wing), the dimensionality method instead focuses on the number of dimensions rather than what those dimensions are. In high-dimensional spaces, we do not need to know exactly what ideologies the dimensions correspond to; the important part is that they signal that there are connections being made between nodes that would not be possible if the network was polarised.

The SVD entropy of the network did not relate to $\hat{d}$ as closely as we expected, though it did reflect major changes in the networks when they occurred. As such, we think it is best to use $\hat{d}$ of the network to measure its polarisation.

A major limitation of this method which could be improved is that it does not capture affective polarisation very well. Our method functionally considers any interaction between two nodes to be “good”; this means that it is not capable of capturing antagonistic interactions between nodes, and as a result it may overestimate $\hat{d}$ of the network by mistaking brief antagonistic reactions for positive social bonds. There is a great deal of scope for integrating the concept of affective polarisation into our model, through methods such as using signed matrices and classification systems such as sentiment analysis to determine whether interactions in a social network are positive, negative, or neutral before determining $\hat{d}$.

Another possible extension of this method would be to implement a nonparametric two-sample hypothesis test\cite{Tang_test_2017}, since this would allow a hypothesis test of whether the two networks are significantly different as additional evidence of polarisation having occurred. We believe that being able to observe the embedded dimensionality of the graph alone is useful; however, we understand that sometimes a hypothesis test is demanded, and we believe this would help demonstrate that changes in the embedded dimensionality of the network are significant.

In our tests using stochastic block models, we found an expected result that $\hat{d}$ increases as the probability of edges forming between the groups increases. This is evidence for the common belief about polarisation that connections between the groups decrease their polarisation. We also found an unexpected result that $\hat{d}$ does not straightforwardly increase when one group is much larger than the other, and in fact decreases greatly when one group is 100 times the size of the other. While this is somewhat a property of the size of the network and its sparsity, since the small group was very small and thus unable to make a large number of edges with the big group, we believe that it is an important case to have tested and that closer investigation of this scenario is warranted. We think that allowing the blocks to become sparsely connected without correcting this property is important for modelling the real conditions of social networks at risk of polarisation, such as communities who have recently received a small number of refugees, and models should be carefully designed to ensure that these effects are retained when appropriate.

It would also be productive to experiment more with stochastic block models, since we only tested two possible scenarios. For example, this paper has only explored the two-block case; many instances of online "echo chambers" have a large number of groups who all hate people different to them, and it would be useful to see what happens to the embedded dimensionality in such cases. Similarly, our testing on group prevalence showed a decrease in embedded dimensionality when one group was 100 times the size of the other, but we did not follow the smaller group’s size all the way to zero so we do not know whether the embedded dimensionality remains low or rebounds to a higher value when one group becomes extremely small.

\section{Acknowledgements}
A preprint version of this article has been peer-reviewed and recommended by PCI Network Sci (https://doi.org/10.24072/pci.networksci.100217) \cite{angst_2025}. We would like to thank them for their excellent process.
 \bibliographystyle{elsarticle-num} 
 \bibliography{cas-refs}

\end{document}